# Numerical simulations of electron tunneling currents in water

Michael Galperin and Abraham Nitzan
School of Chemistry, Tel Aviv University, Tel Aviv, 69978, ISRAEL
and

Ilan Benjamin
Department of Chemistry, University of California, Santa Cruz, CA 95064, USA

## Abstract

This paper presents results of numerical simulations of electron tunneling through water that extend our previous calculations on such systems in several ways. First, a tip-substrate configuration is used; second, calculations are carried in the presence of an external potential bias; third, the image potential that reflects the interaction of the electron with the mobile metal electrons is taken into account in the static image approximation. Finally, all-to-all transmission probability calculations are performed in order to get an order-of-magnitude estimate of the current-voltage characteristics of this junction model. The computed currents are within the range of the few available experimental observations on scanning tunneling microscope (STM) currents in water, indicating that our calculation may have taken into account all the important physical attributes of such systems. In addition we examine the effect of the water medium on the spatial distribution of the tunneling flux. We find that while different water configurations scatter the tunneling electron in different ways, on the average the water affected loss of resolution is rather small in the deep tunneling regime but can be substantial in energy regimes where the tunneling is strongly affected by water-supported resonance structures.

## I. Introduction

Electron tunneling through water is an important element in all electron transfer processes involving hydrated solutes, and in many processes that occur in water based electrochemistry. Only a few systematic experimental studies of the effect of the water structure on electron transfer processes have been done.[1-13] Early theoretical treatments have modeled water as a dielectric continuum, however there are indications that the discrete 3-dimensional structure of the water environment affects the way tunneling takes place. Porter and Zinn[3] have found, for a tunnel junction made of a water film confined between two mercury droplets, that at low (<1nm) film thickness conduction reflects the discrete nature of the water structure. Nagy [9-11] have studied STM current through adsorbed water layers and has found that the distance dependence of the tunneling current depends on the nature of the substrate and possibly indicates the existing of resonance states of the excess electron in the water layer. Vaught et al[8] have seen a non-exponential dependence on tip-substrate distance of tunneling in water, again indicating that at small distances water structure and possibly resonance states become important in affecting the junction conductance. Several workers have found that the barrier to tunneling through water is significantly lower than in vacuum for the same junction geometry.[1,2,14,4,11,7,12,13] The observed barrier is considerably lower than the threshold observed in photoemission into water[15,16] and, while image effects should be taken into account when the tunneling takes place near an electrode surface, their role in less obvious than in the corresponding process in vacuum.[3] In several recent papers we [17-26] and others[27,28] have attempted to correlate these observations with numerical and theoretical studies. Our studies have indicated that electron tunneling is indeed strongly affected by the 3-dimensional structure of the water layer. In particular we have identified several sources that affect the apparent barrier to tunneling through water as compared with vacuum. First, the occupation of a substantial fraction of the space between the electrodes by the cores of oxygen atoms, which are essentially impenetrable to the tunneling electron contributes to increase the apparent barrier. Second, the ordering of water molecule on the metal surface is often such that causes reduction of the metal work-function and consequently a lower barrier to tunneling. Finally, tunneling is enhanced by transient resonances supported by cavities in the water structures. Figure 1 shows a compilation of numerical results obtained for 20 water configurations sampled from an equilibrium trajectory (300K) of water between two parallel Pt(100) planes separated by 10Å. This space contains three water monolayers at density 1gr/cm$^3$. Shown is the total transmission probability as a function of incident kinetic energy for an electron plane wave incident normal to the Pt surface. (For more details of the model, the potentials used and the numerical method see Refs. 19,20,23). The vacuum barrier in this calculation was 5eV, and the structure in the transmission in the range of ~1eV below it arises from the aforementioned resonances. The distribution of the corresponding peaks reflects the transient nature of these resonances that depend on the evolving water structure. In addition, it is seen that the existence of these resonances also enhances the transmission probability in the deep tunneling regime.

These calculations where done using static water structures confined between two parallel planar electrodes in the absence of potential bias. Image effects were not considered explicitly - it was assumed that the vacuum potential used already incorporates the image contribution that, because of the given geometry, does not depend on the lateral coordinates. In most experimental situations however a tip with assumed atomic dimensions is used as one of the electrode and the current is monitored as a function of imposed voltage. Also, a calculation of the type displayed in Fig. 1, which is based on a given incoming direction has to be summed over all incoming directions in order to evaluate the current. The present paper describes calculations of electron tunneling through water that take into account these needed generalizations. We note in passing that another generalization that takes into account the water nuclear dynamics and the associated inelastic contribution to the tunneling current has been presented in a separate publication.[26] We have found that inelastic tunneling contributions are substantial, however they are dominated by low energy phonons and therefore do not affect appreciably the electron energy in what might be termed a quasi-elastic process. The present study, like earlier ones,[23] was carried out using a collection of static equilibrium water configurations.

The next Section briefly describes our model and method of calculation. In Sect. 3 we present results for tunneling in an ideal "STM-configuration" - water filling the gap between an atomic tip and an ideal substrate surface. Section 4 summarizes our findings.

**2. Model and computational method**

A projection of the STM-like junction used in our calculation is shown in Fig. 2 The shaded gray areas are metal electrodes, with the planar surfaces corresponding to the (100) surface of Pt. The protrusion on the left electrode represents a tip whose structure in the present calculations is described below. In the following discussion the tip point is taken to be the origin of a cartesian coordinate system with the z axis going through the tip axis as shown. A potential bias, $\Delta\Phi = \Phi_2 - \Phi_1$ is imposed and the steady state current in monitored as a function of this bias. $\Phi_1$ and $\Phi_2$ are assumed constants on the corresponding electrode surfaces. The space between the electrodes is filled with water whose density was taken to be $1g/cm^3$. Molecular dynamics at 300K is used to generate equilibrium water configurations in the space between the electrodes. In this calculation we use a cell of lateral dimensions $L_x=L_y=39.242$Å with periodic boundary conditions. The water potential was taken to be the polarizable flexible simple point charge model (PFSPC) described in Refs. 22,29, and the water-Pt interaction was taken from Refs. 30,31. In addition the water is subjected to the external field associated with the imposed bias. To find this force we use the Laplace equation

$$\nabla^2 \Phi = 0 \tag{1}$$

with periodic boundary conditions in the x-y direction and Dirichlet boundary conditions $\Phi=\Phi_1$ and $\Phi=\Phi_2$ on the left and right electrode surfaces, $S_1$ and $S_2$, respectively.

In the bulk of the metal electrodes the electron is assumed to behave as a free particle of mass $m=m_e$. Next we construct the potential for the electron motion in the space between the electrodes. We approximate this potential by the sum

$$V = V_B + V_W + V_E + V_I \tag{2}$$

where $V_B$ is the bare potential barrier whose height is given by the difference between the vacuum potential and the bottom of the metal conduction band, $V_W$ is the electron water pseudo-potential, taken from the work of Barnett et al and modified to include the effect of the water electronic polarizability (for more details see Refs. 20,23), $V_E$ is the potential energy associated with the external voltage, $V_E = -e\Phi$, where $\Phi$ is the solution of Eq. (1) and, finally, $V_I$ is the potential associated with the interaction of the excess electron in the barrier with the metal charge distribution, i.e. the image potential. The latter is an enigma in this type of calculation as is the way the water electronic polarizability is included, since their inclusion is just a poor man's way to account for many-electron correlations in the process. Here we follow past works on electron tunneling near metal surfaces, as well as recent literature on image states at such surfaces and represent the image interaction by its static limit (i.e. we assume that the timescale associated with a tunneling event is long relative to the response time of the metal electron (measured by the metal plasma frequency). In this static limit we approximate $V_I(\mathbf{r})$ by the classical image potential for a point particle of charge e at position $\mathbf{r}$. Even then, the determination of the image contribution is subject to inaccuracies resulting from the uncertainty about the location of the image plane and the way the singularity at the metal surface is handled, and we have arbitrarily chosen to truncate the potential at the point where it reaches the bottom $E_0$ of the metal conduction band. Near a planar surface this would mean

$$V_I(\mathbf{r}) = \begin{cases} -e^2/4z & \text{for } z > z_0 \\ -e^2/4z_0 & \text{for } z \leq z_0 \end{cases} \quad (z_0 \text{ defined from } e^2/4z_0 = E_0) \qquad (3)$$

where $z$ is the distance from the metal plane. Between the two metal surfaces of Fig. 2, in particular in the presence of the tipped surface, $V_I(\mathbf{r})$ has to be evaluated numerically. The numerical procedure that accomplishes this task is described elsewhere.[32] This calculation is based on solving a Poisson equation for each position of the electron under the boundary conditions determined by the electrode surfaces, and subtracting the singularity associated with the self-interaction of the electron using analytical results for a point charge between two planar surfaces. In the calculations described below we have used $E_0 = -12.4\text{eV}$ (measured from vacuum) and $E_F = 7.3\text{eV}$ (above $E_0$, i.e. workfunction=5.1eV).

Results of these electrostatic calculations are shown in Figure 3. Fig. 3a shows a $y=0$ cut of the external potential $\Phi$ between the surfaces displayed in fig. 2. The tip is taken as a pyramid constructed by connecting the centers of atoms on the surface of the atomic tip described below. The base of this pyramid is a square of dimension $5a/\sqrt{2}$ and its height is $2.5a$ where $a=3.9242\text{Å}$ is the Pt lattice constant, and the distance of the tip point from the opposite surface is 9Å. Fig. 3b displays the image potential computed as described above and plotted as function of the distance between the two metals along the $z$ axis, for $y=0$ and for $x=0$ (a line going through the tip axis), $x=11.96$au and $x=23.92$au. The fact that the height of these potential sections do not vary much with the effective distance between the metal surfaces (as compared e.g. with the results obtained between two planar surfaces) reflects the 3-dimensional tip structure of the present junction.

We note in passing the imposition of a potential bias between the electrodes affects the electron tunneling both directly and indirectly through its effect on the water structure. Fig. 4. shows the effect of a bias of 0.5V on the water structure as expressed by the angular distribution of the molecular dipole direction relative to the tip axis (i.e. the direction normal to the flat electrode). In Fig. 4a only those molecules whose oxygen centers are located within a cylinder of radius R=2Å about the tip axis are included. In Figs 4b and 4c this cylinder is taken with R=6Å and 12Å respectively. It is seen that field induced structural effects are most important in the vicinity of the tip. They seem to be negligible far enough from it, however they are always important for the first monolayers directly at the electrodes surfaces.[17] The effect of preferential water ordering on electron tunneling has been studied in Ref. 21.

In the work described in Refs. 17-21,23,25-26,33 we have evaluated the transmission probability through the given potential barrier using the absorbing boundary condition (ABC) Green's function technique of Seideman and Miller.[34,35] The Hamiltonian is represented on a spatial Grid in the range $-L_x/2 < x < L_x/2; -L_y/2 < y < L_y/2; -L_z/2 < z < L_z/2$ with periodic boundary conditions in the *xy* plane and absorbing boundary conditions near the system edges along *z*. Absorption at these boundaries is affected by suitable imaginary potentials, $\varepsilon_L(z)$ and $\varepsilon_R(z)$ that smoothly increase from zero towards the system edges along the z axis. For a particle of energy E incident on the barrier from the left (say) the "one-to-all" (i.e. a given incident direction and a sum over all final directions) transmission probability is given by

$$T[\phi_{in}(E)] = \frac{2}{\hbar} <\phi_{in}(E) | \varepsilon_L G(E) \varepsilon_R G(E) \varepsilon_L | \phi_{in}(E)> \qquad (4)$$

while the "all-to-all" transmission probability (that includes a sum also over all incident directions) is obtained from

$$T(E) = 4 Tr\left[ G(E) \varepsilon_R G^\dagger(E) \varepsilon_L \right] \qquad (5)$$

In these expressions *G* is the Greens function of the system Hamiltonian supplemented by the absorbing potential, i.e.,

$$G(E;\varepsilon) = [E - H + i\varepsilon(\mathbf{r})]^{-1} \quad ; \quad \varepsilon(\mathbf{r}) = \varepsilon_L(\mathbf{r}) + \varepsilon_R(\mathbf{r}) \qquad (6)$$

and the absorbing potentials $\varepsilon_L(z)$ and $\varepsilon_R(z)$ are chosen arbitrarily provided two conditions are satisfied: First, these functions should vanish in the interior system where the actual scattering takes place. Second, these potentials should rise smoothly towards the edges of the system in the tunneling or scattering directions so as to insure a full absorption of all outgoing waves. For more information on the choice of these functions in the present context see Refs. 17,24.

We have recently shown that Eqs. (4) and (5) are in fact approximations to the exact expressions obtained when a quantum problem in infinite space is projected onto the subspace in which the actual interactions occur, for example

$$\mathcal{T}(E) = Tr\left[G(E)\Gamma_R(E)G^\dagger(E)\Gamma_L(E)\right] \quad (7)$$

in which G is, as before, the Green's function of the of the given subspace

$$G(E) = \left(E - H - \Sigma_L(E) - \Sigma_R(E)\right)^{-1} \quad (8)$$

where $\Sigma(E)$ is the self energy associated with the truncation procedure, that carries the effect of the rest of the universe on this subspace, and where $\Gamma(E) = -2\,\text{Im}(\Sigma(E))$. If $\Sigma(E)$ was known exactly Eq. (7)-(8) would provide a way for the exact evaluation of the transmission probability. The ABC Green's function methodology is based on the observation that if the system boundaries are taken far enough from the scattering region the exact self-energy is unimportant as long as the condition of absolute absorbance at the system boundary is satisfied. This is the basis for replacing the (usually non-local) exact self energy by the local functions $\varepsilon_L(z)$ and $\varepsilon_R(z)$. We note that the need to place the system's boundaries far enough from the scattering region implies the need to use a very large basis set, i.e. a very large grid, to describe the tunneling process. For this reason we were able in Refs. 18-21,23-24 to calculate only one-to-all transmission probabilities, while all-to-all calculations needed to evaluate the overall tunneling current where too demanding given our computing resources.

On the other hand, if the exact self energy is known, the system boundaries in the tunneling (z) direction could be placed just outside the scattering region, e.g. on surfaces $S_1$ and $S_2$ (see Fig. 2), implying a substantially smaller system described in terms of a smaller number of grid points. This goal in fact can be achieved. In the spatial grid representation, using a finite difference approximation for the kinetic energy operator, a free particle Hamiltonian has a tight binding structure, in which case the self-energy associated with truncating the system on any planar boundary can be computed exactly. The detailed procedure for carrying out such a program is described elsewhere.[32] In a typical calculation this makes it possible now to use a system whose dimension in the tunneling direction is ~1/3 the size used in our previous work, implying an order of magnitude reduction in cpu time. This makes it possible for us to compute for the first time all-to-all transmission probabilities in the tip-water-substrate junctions of the kind modeled by Fig. 2.

Results obtained using the methodologies outlined above are presented next. The calculation reported were done on a model system in which two parallel Pt (an FCC solid with a lattice constant $a$=3.9242Å and nearest neighbor atom distance $a/\sqrt{2}$) slabs are placed with spacing D (along the z axis) between them, so that the surface planes (001) are facing each other. The size of each slab is $10a$ in the lateral directions x and y and $1.5a$ in the z direction (the latter corresponds to 3 (001) monolayers of Pt atoms in the FCC lattice. The tip is constructed by placing five additional (001) layers on the left slab with decreasing number of atoms: 25, 16, 9, 4, 1, where the upper layer containing one atom constitutes the tip point, so that the distance between this point and the opposite (substrate) surface is $d = D - 2.5a$.

## 3. Numerical Results

The present calculations generalize our previous work in four important ways. First a tip structure is included making it possible to study realistic scanning tunneling microscope configurations. Secondly, a potential bias is imposed between the tip and substrate electrodes. Third, image effects are taken into account, albeit in a crude approximate way. Finally all-to-all transmission calculations are done making it possible to estimate theoretically the current in aqueous tunneling junctions. Two groups of numerical results are shown and discussed below. In the first we focus mainly on the presence of the tip in the tunneling junction, and discuss the interplay between the focusing effect of the tip and the dispersion caused by the water. In the second we combine the ingredients outlined above to compute current vs. voltage in our model of underwater STM experiment.

Given the junction configuration of Fig. 2, tunneling in a uniform medium (e.g. vacuum) is expected to be dominated by the one-dimensional path running along the tip axis and normal to the planar electrode. Fig. 5a shows the distribution of the $z$ component of the tunneling flux, $J_z$, obtained for this case, for tunneling with incident energy 1.38eV below the vacuum barrier.[1] The electrodes are not shown on these plots, except that a projection of the pyramidal tip is shown as a reference. The electron is incident as a plane wave in the bulk of the left electrode in the positive z direction and the tunneling flux distribution in the space between the electrodes is obtained from

$$\mathbf{J} = \frac{\hbar}{2mi}\left(\psi^* \cdot \nabla\psi - \psi \cdot \nabla\psi^*\right) \tag{9}$$

using the exact scattering wavefunction computed as described in Ref. 24. This flux is shown by forming contour plots (neighboring contours differ by a fixed flux increment) on consecutive *xy* planes and different values of *z* and projecting them all on the plane of the paper. The inset shows the flux distribution as a function of *x* for *y*=0 and for *z*=2.5Å and 5Å, both normalized to 1 at their maximum. It is seen that while the flux has some spatial distribution, it is sharply peaked about the (0,0,*z*) axis as expected.

The situation can be quite different in the presence of water, and depends of course on the water configuration. Two distinct scenarios are displayed in Figs. 5b and 5c. (1.38eV below vacuum in Fig. 5b; 1.27 eV below vacuum in Fig. 5c). Fig. 5b shows a situation with no preferred path, and the tunneling electron seems to be scattered strongly in a large part of available space. In contrast, Fig. 5c shows a situation where a resonance supporting structure exists at some location inside the water barrier. The flux distribution indicates that the electron's preferred path goes through this resonance even though this means a strong deviation from the shortest straight-line path. Obviously, these different characteristics are properties of individual water configurations and will not be seen after time or ensemble averaging.

These observations of the behavior of the flux distribution are potentially relevant to assess the influence of the water medium on the spatial resolution of an underwater STM image. As an example we show in Fig. 6 the root of the second moment of $J_z$ on the surface of the planar electrode as a function of its distance from the tip. This second moment is defined by

$$M_2(z) = \int dx \int dy \left(x^2 + y^2\right) J_z(x,y,z) \Big/ \int dx \int dy J_z(x,y,z) \tag{10}$$

---

[1] The results depicted in Fig. 5 were obtained with the same choice of barrier parameters as in Refs. 17-26., i.e. a vacuum barrier represented by a rectangular potential of height 5eV. Image effects are not included.

Fig. 6a displays this measure of the width of the electron distribution on the substrate surface as a function of substrate-tip distance for the case where the incident electron energy is 2eV below the vacuum barrier. Shown are results for a rectangular ("vacuum") barrier, for a similar junction with the vacuum replaced by water (density 1g/cm$^3$), and for the same barrier where the water is replaced by a random distribution of hard spheres of the same density and radius equal to that of the oxygen atom. Note that the result shown for the water barrier represent an average over 10 equilibrium (300K) water configurations. Fig. 6b shows similar results for electron energy 1.3eV below the vacuum barrier. Two observations are immediately evident: first, a distribution of hard spheres hardly affects the width of the electron current distribution over the lengthscale considered. Second, the presence of water in the space between the electrodes does affect the tunneling flux distribution, however this effect is far greater in the energy regime closer to the vacuum barrier where tunneling is affected by barrier resonances, while the effect in the deep tunneling regime is modest.

We now turn to current vs. voltage calculations. For this purpose we need to prepare equilibrium water configurations under a given potential bias between the electrodes, evaluate the different contributions to the barrier potential, Eq. (2), then calculate all-to-all transmission probabilities as function of electron energy for the given potential bias. The tunneling current for this voltage $\Delta\Phi$ is then computed from[36]

$$I = \frac{e}{\pi\hbar} \int_0^\infty dE \left[ f_L(E) - f_R(E + e\Delta\Phi) \right] \mathcal{T}(E) \qquad (11)$$

It should be emphasized that since tunneling processes are extremely sensitive to details of the potential barrier, we should not hope to reproduce exact experimental observations of STM current-voltage characteristics. In addition, our computing resources are not sufficient to carry tunneling calculation on the many water configurations needed for a good statistical average. Rather, our aim is to test whether our model calculation yields a reasonable order of magnitude estimate for electron transport in tip-water-substrate junctions and to re-examine the issue of the low effective barrier to tunneling in such systems.

Results of these calculations are summarized in Figure 7a. The top and bottom lines represent the current against bias voltage between tip and substrate separated by 5.85Å (2 water monolayers) and 12.15Å (4 water monolayers). These are results obtained from single water configurations. The intermediate group of lines represent similar results obtained for 5 different water configurations at tip-substrate separation 9Å, corresponding to 3 water monolayers. Fig. 7b shows the average of these five results as well as a linear least square fit. These results obviously suffer from insufficient statistical averaging. Still, the following observations can be made:

(a) The order of magnitude of the computed current, order of 0.1nA at $\Delta\Phi$=0.5V at 9Å tip-substrate separation is within reasonable agreement with the (scattered) available experimental data, which are scattered in the range 0.1-1nA for this voltage and distance range.[5,6,8,13] Given the uncertainties discussed above concerning the electron-water potential, the tip structure and the evaluation of the image potential, and given our poor statistics this is as much as we can hope for in this type of calculation.

(b) The presence of water in the space between the electrodes causes a substantial decrease in the effective junction barrier. Fitting the currents computed in Fig. 6a to similar calculations done on a rectangular barrier model of a similar width, we find effective barrier heights of 1.7, 2.0 and 2.8eV (energies above the Fermi energy) for the tip substrate distances of 5.85, 9.0 and 12.15Å, respectively. Again, these values are of the order estimated from experimental data. It should be noted that the dependence of the effective barrier height on the tip-substrate separation reflects the structure of the image potential. Indeed, the top of the image potential curves displayed in Fig. 3b are 2.24, 3.14 and 3.75eV respectively. For this reason, fitting the distance dependence data of Fig. 7a to a simple exponential dependence $I \sim e^{-\kappa d}$ with a constant $\kappa$, does not reflect this more complex situation.

## 4. Summary and conclusions

We have performed simulations of electron tunneling through water that extend our previous calculations on such systems in several ways. First, a tip-substrate configuration is used; second, calculations were carried in the presence of an external potential bias; third, the image potential that reflects the interaction of the electron with the mobile metal electrons is taken into account in the static image approximation. Finally, all-to-all transmission probability calculations were performed in order to get an order-of-magnitude estimate of the current-voltage characteristics of this junction model.

Both the external bias and the image interactions are obviously necessary ingredients in any calculation of electronic currents in molecular junctions. Relative to our previous calculations that have disregarded these factors, the lowering of the barrier by the image interaction seems to be the most important. This also has the effect of broadening the resonance structure found in our earlier studies,[24] however the contribution of these resonances to the further average reduction of the effective barrier is not expected to change. Other effects arise from the dependence of the water structure on the bias field. We have shown before[21] that this can be an important effect in large fields, but in the moderate fields considered in this paper (and normally used in the laboratory) it seems to be within the noise associated with the smallness of our sample and cannot be assessed with our limited statistics.

The currents computed in this work are within the range of the few available experimental observations, indicating that our calculation may have taken into account all important physical attributes of this systems. In addition we have examined the effect of the water medium on the spatial distribution of the tunneling flux. We have found that while different water configurations scatter the tunneling electron in different ways, on the average the water affected loss of resolution is rather small in the deep tunneling regime but can be substantial in energy regimes where the tunneling is strongly affected by water-supported resonance structures. Together with our earlier results of inelastic effects in tunneling through water[26], these studies provide a complete qualitative picture of electron tunneling through this important medium. It should be kept in mind though that several important questions still remain unanswered. For example, our poor statistics makes it impossible to analyze at present the quantitative effect of water and of the existence of tunneling resonances on the image resolution in underwater STM studies. Similarly, a quantitative assessment of the non-linear current-voltage behavior expected from the field effect on the water structure awaits a more extensive numerical work. Finally, the possible role of rare structural fluctuations in the deep tunneling regime[24] is an important subject for future study.

**Acknowledgements**. This work is dedicated to Professor Steve Berry on his 70th birthday for his fundamental contributions to chemistry research. This research was supported in part by the U.S.A.-Israel Binational Science Foundation. The research of AN is also supported by the Israel Ministry of Science, and the research of IB is also supported by a grant from the National Science Foundation (CHE-9981827).

**Figure captions**

Figure 1. A compilation of numerical results for the transmission probability as a function of incident electron energy, obtained for 20 water configurations sampled from an equilibrium trajectory (300K) of water between two planar parallel Pt(100) planes separated by 10Å. The vacuum is 5eV and the resonance structure seen in the range of 1eV below it varies strongly between any two configurations. Image potential effects are disregarded in this calculation.

Figure 2. A projection of the model junction used in the calculations presented in the Figs. 3-7. See text for details.

Figure 3. Results of computed electrostatic contributions to the single electron potential in the junction. (a) A y=0 cut of the external potential distribution between the tip and the flat substrate (that corresponds to the $V_E$ term in Eq. (2)) for a voltage drop of 0.5V between these electrodes. (b) The image potential computed along different lines normal to the flat electrodes: (1) x=0 (a line going through the tip axis); (2) x=11.96au (distance from the tip axis); (3) x=23.92au.

Figure 4. The angular distribution of water molecules in the junction (Fig. 2) under a bias of 0.5V at T=300K (solid lines). (a) Molecules located within a cylinder of radius R=2Å, whose axis is the line normal the flat electrode going through the tip. (b) Same with R=6Å. (c) Same with R=12Å. In these calculations the vacuum barrier was taken 5eV and the separation between the tip and the flat electrode was 9Å. The dashed lines are the corresponding results in the absence of a potential bias.

Figure 5. The spatial distribution of $J_z$, the *z* component of the tunneling flux, displayed as a contour plot taken on successive *xy* planes along the tunneling direction z. (a) Vacuum. The inset shows the flux distribution as a function of *x* for *y*=0 and for *z*=2.5Å (full line) and 5Å (dotted line), i.e. at 2.5Å and at 5Å to the right of the tip point in Fig. 2, both normalized to 1 at their maximum. (The ratio between the absolute flux values at these maxima was 13.6). (b)-(c) Water. See text for details. The junction parameters are as in Fig. 4.

Figure 6. The spread of the *z* component of the tunneling flux distribution about the direction defined by the tip axis, expressed by the square root of the second moment of this distribution calculated on the substrate surface ($M_2(z=d)$) from Eq. (10), plotted against the tip-substrate distance d. The three lines show results obtained from different inter-electrode media: Dashed line with squares - vacuum. Dotted line with triangles - a random distribution of hard spheres of radius and number density same as the water oxygen cores. Full line with circles - one configuration of water. The vacuum barrier is 5eV. (a) Incident electron energy is 2eV below the vacuum barrier ("deep tunneling"). (b) Incident energy 1.3eV below the vacuum barrier (closer to the resonance tunneling regime).

Figure 7. Current against bias voltage obtained from all-to-all tunneling calculations in water, in a biased tip-planar electrode junction. (a) The upper and lower lines are results of calculation done on single configurations characterized by tip-substrate separation of 5.85Å (2 water monolayers) and 12.15Å (4 water monolayers), respectively. The intermediate group of lines represent similar results obtained from 5 different water configurations at tip-substrate separation 9Å, corresponding to 3 water monolayers. (b) The average of the five results obtained at tip-substrate separation 9Å, together with a linear least square fit to these results.

**REFERENCES**

(1) Bingelli, M.; Carnal, D.; Nyffenegger, R.; Siegenthaler, H.; Christoph, R.; Rohrer, H. *J. Vac Sci Technol. B* **1991**, *9*, **1985.**

(2) Haiss, W.; Lackey, D.; Sass, J. K.; Besocke, K. H. *J. Chem. Phys.* **1991**, *95*, **2193.**

(3) Porter, J. D.; Zinn, A. S. *J. Phys. Chem.* **1993**, *97*, **1190-1203.**

(4) Meepagala, S. C. *Phys. Rev. B* **1994**, *49*, **10761.**

(5) Halbritter, J.; Repphun, G.; Vinzelberg, S.; Staikov, G.; Lorentz, W. J. *Electrocimica Acta* **1995**, *40*, **1385-1394.**

(6) Repphun, G.; Halbritter, J. *J. Vac. Sci. Technol.* **1995**, *13*, **1693-1698.**

(7) Pan, J.; Jing, T. W.; Lindsay, S. M. *Chem. Phys.* **1994**, *98*, **4205.**

(8) Vaught, A.; Jing, T. W.; Lindsay, S. M. *Chem. Phys. Lett.* **1995**, *236*, **306-310.**

(9) Nagy, G. *J. Electroanal. Chem.* **1995.**

(10) Nagy, G. *Electrochimica Acta* **1995**, *40*, **1417-1420.**

(11) Nagy, G. *J. Electroanal. Chem.* **1996**, *409*, **19-23.**

(12) Hahn, J. R.; Hong, Y. A.; Kang, H. *Applied Physics A-Materials Science & Processing* **1998**, *66*, **S467-S472.**

(13) Hong, Y. A.; Hahn, J. R.; Kang, H. *Journal of Chemical Physics* **1998**, *108*, **4367-4370.**

(14) Hamin, H. J.; Ganz, E.; Abraham, D. W.; Thomson, R. E.; Clarke, J. *Phys. Rev. B* **1986**, *34*, **9015.**

(15) Gurevich, Y. Y.; Pleksov, Y. Y.; Rotenberg, Z. A. *Photo-electrochemistry*; Consultant Bureau: New York, 1980.

(16) Konovalov, V. V.; Raitsimring, A. M.; Tsvetkov, Y. D. *Radiat. Phys. Chem.* **1988**, *32*, **623.**

(17) Mosyak, A.; Nitzan, A.; Kosloff, R. *Journal of Chemical Physics* **1996**, *104*, **1549-1559.**

(18) Evans, D.; Benjamin, I.; Seideman, T.; Ezer, H. T.; Nitzan, A. *Abstracts of Papers of the American Chemical Society* **1996**, *212*, **194-COLL.**

(19) Benjamin, I.; Evans, D.; Nitzan, A. *J. Chem. Phys.* **1997**, *106*, **6647-6654.**

(20) Mosyak, A.; Graf, P.; Benjamin, I.; Nitzan, A. *Journal of Physical Chemistry a* **1997**, *101*, **429-433.**

(21) Benjamin, I.; Evans, D.; Nitzan, A. *J. Chem. Phys.* **1997**, *106*, 1291-1293.

(22) Benjamin, I. *Chem. Phys. Lett.* **1998**, *287*, 480.

(23) Nitzan, A.; Benjamin, I. *Accounts of Chemical Research* **1999**, *32*, 854-861.

(24) Peskin, U.; Edlund, A.; Bar-On, I.; Galperin, M.; Nitzan, A. *J. Chem. Phys.* **1999**, *111*, 7558-7566.

(25) Galperin, M.; Nitzan, A.; Peskin, U. *to be published* **2001**.

(26) Galperin, M.; Nitzan, A. *J. Chem. Phys.* **2001**, *115*, 2681-2694.

(27) Schmickler, W. *Surface Science* **1995**, *335*, 416-421.

(28) Peskin, U.; Edlund, A.; Bar-On, I. *Journal of Chemical Physics* **2000**, *112*, p.3220-6.

(29) Dang, L. X. *J. Chem. Phys.* **1992**, *97*, 2659-2660.

(30) Spohr, E. *Phys. Chem.* **1989**, *93*, 6171.

(31) Spohr, E.; Heinzinger, K. *Ber. Bunsen-Ges. Phys. Chem.* **1988**, *92*, 1358.

(32) Galperin, M.; Toledo, S.; Nitzan, A. *to be published* **2002**.

(33) Nitzan, A. *Ann. Rev. Phys. Chem.* **2001**, *52*, 681- 750.

(34) Seideman, T.; Miller, W. H. *J. Chem. Phys* **1992**, *97*, 2499.

(35) Seideman, T.; Miller, W. H. *J. Chem. Phys* **1992**, *96*, 4412.

(36) See, e.g. Datta, S. *Electric transport in Mesoscopic Systems*; Cambridge University Press: Cambridge, 1995.

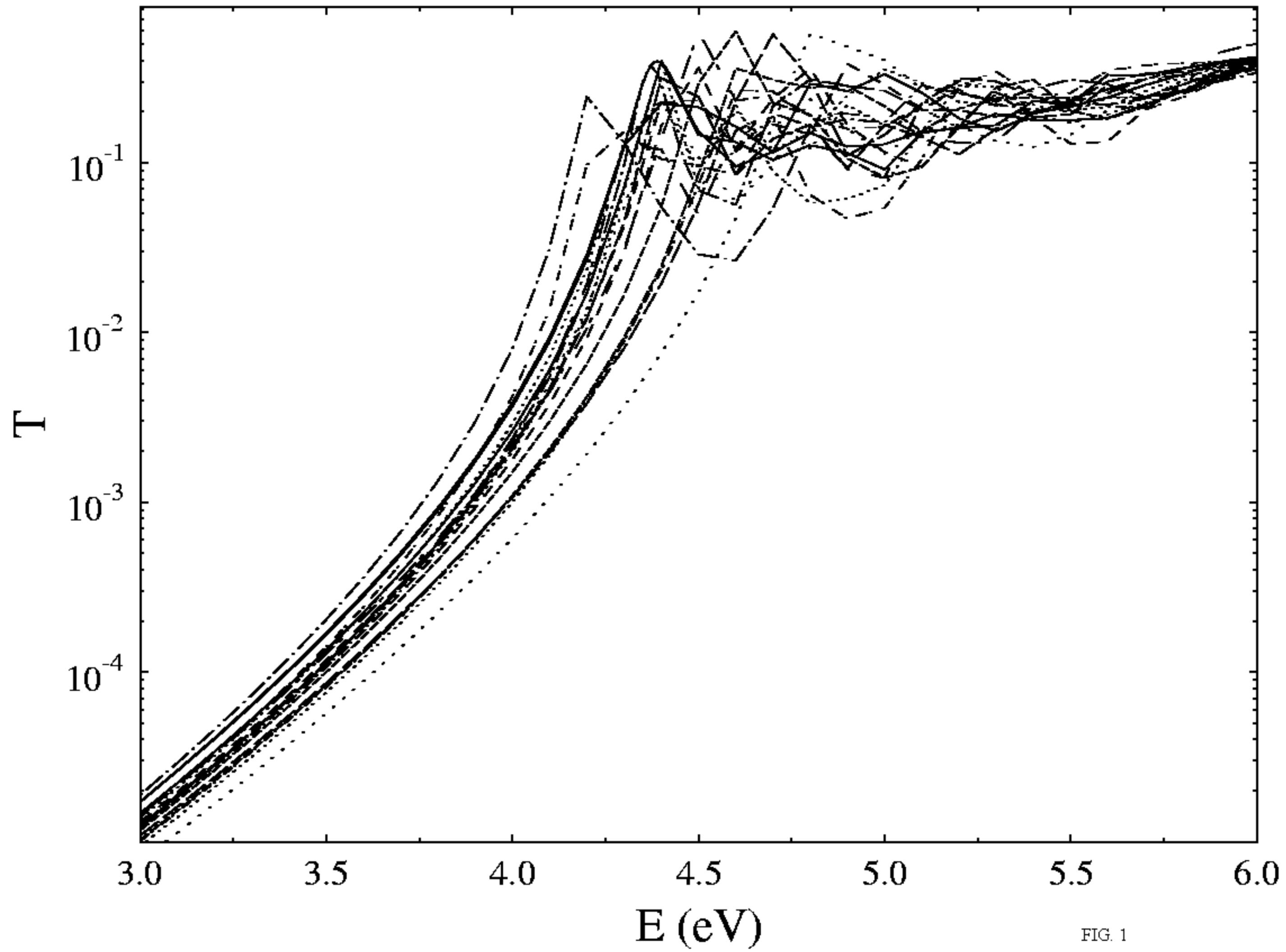

FIG. 1

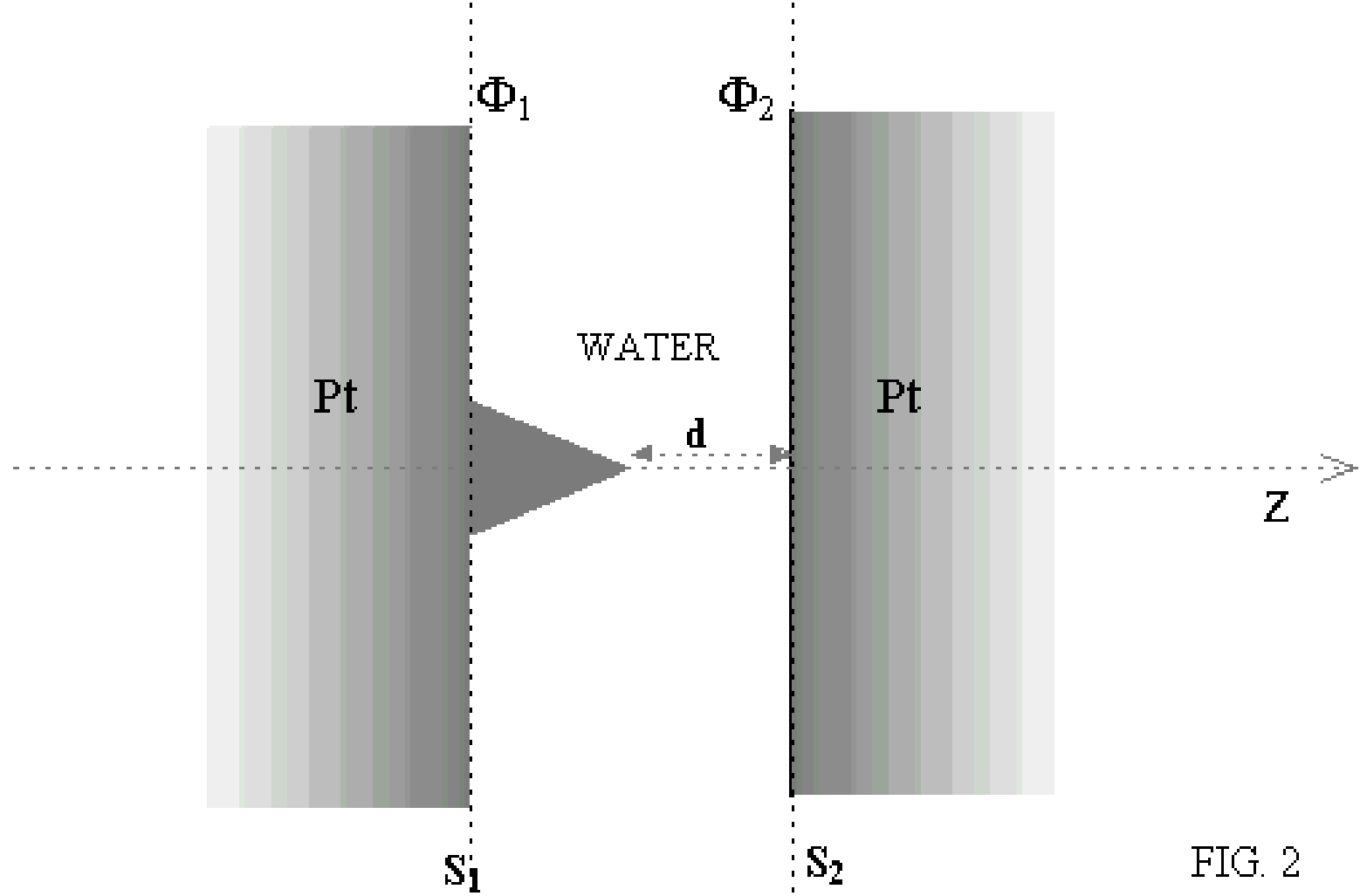
FIG. 2

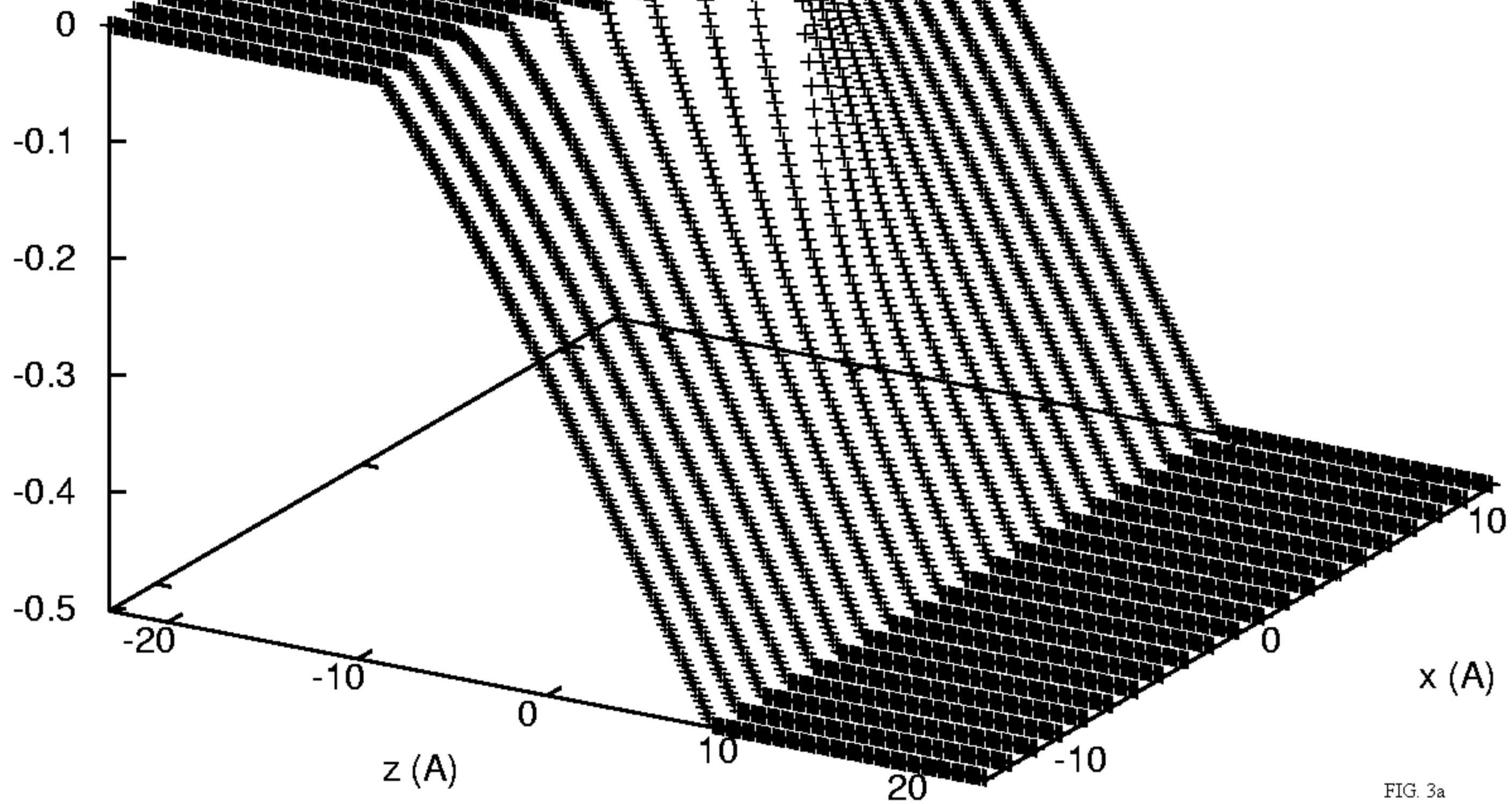

FIG. 3a

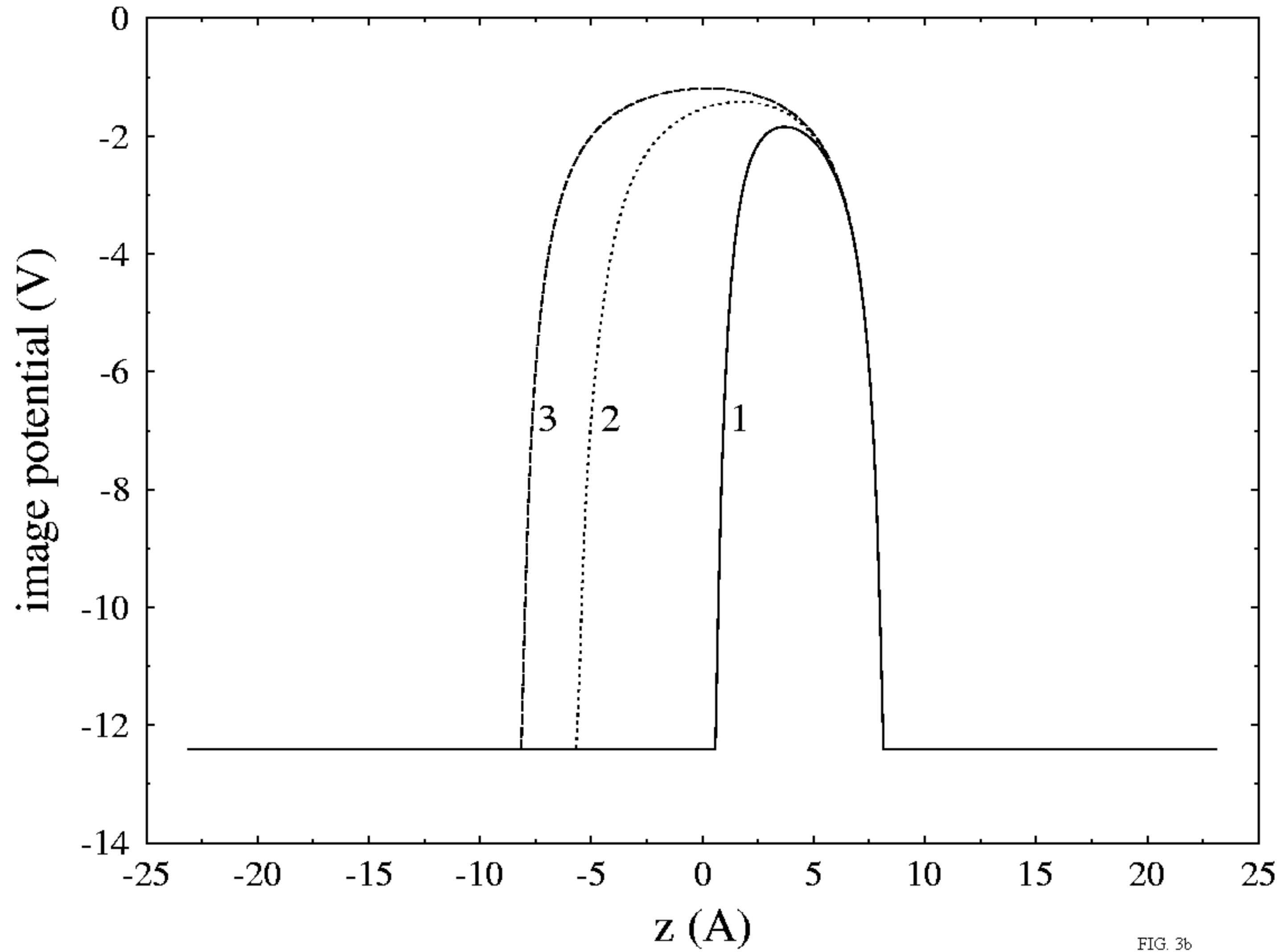

FIG. 3b

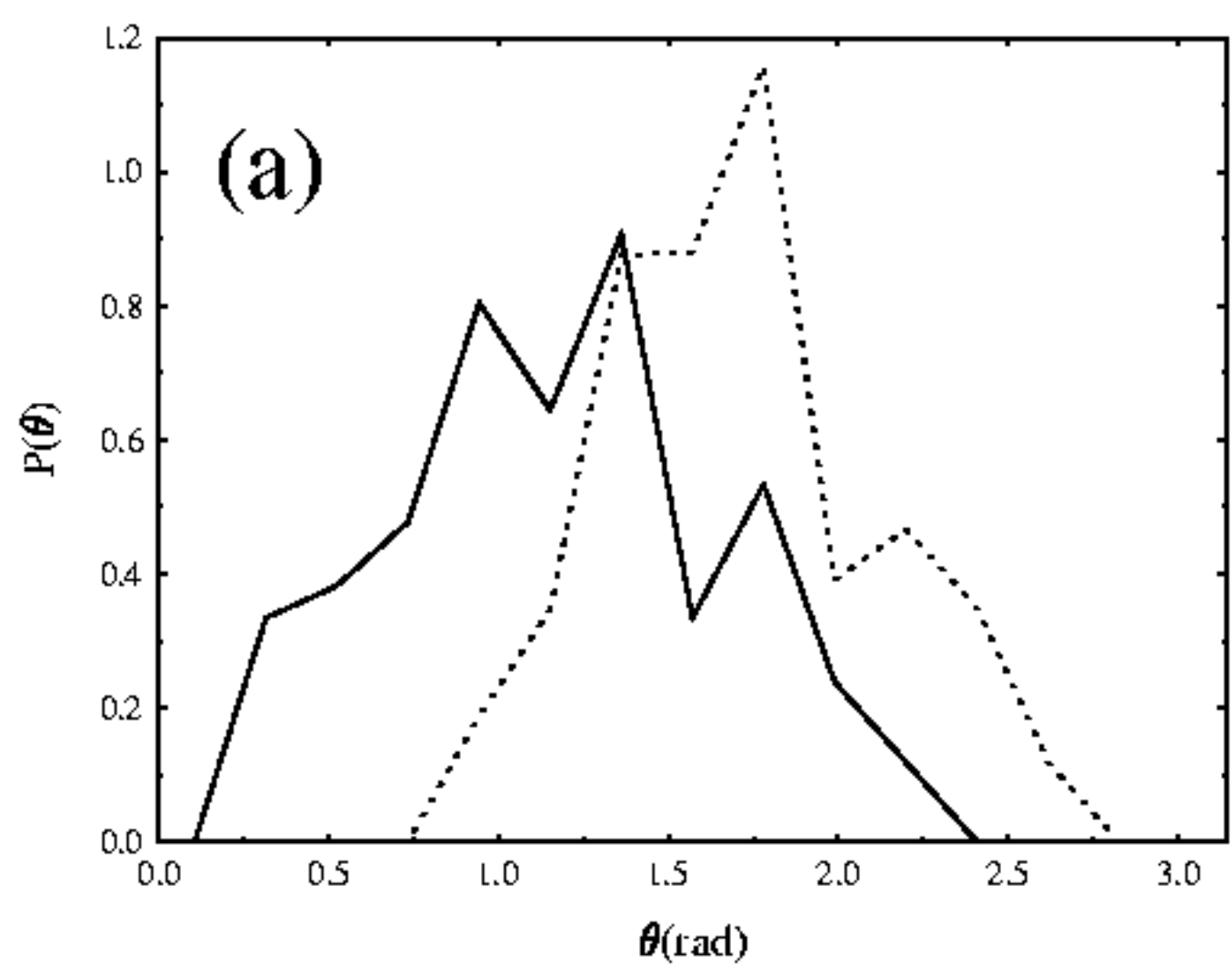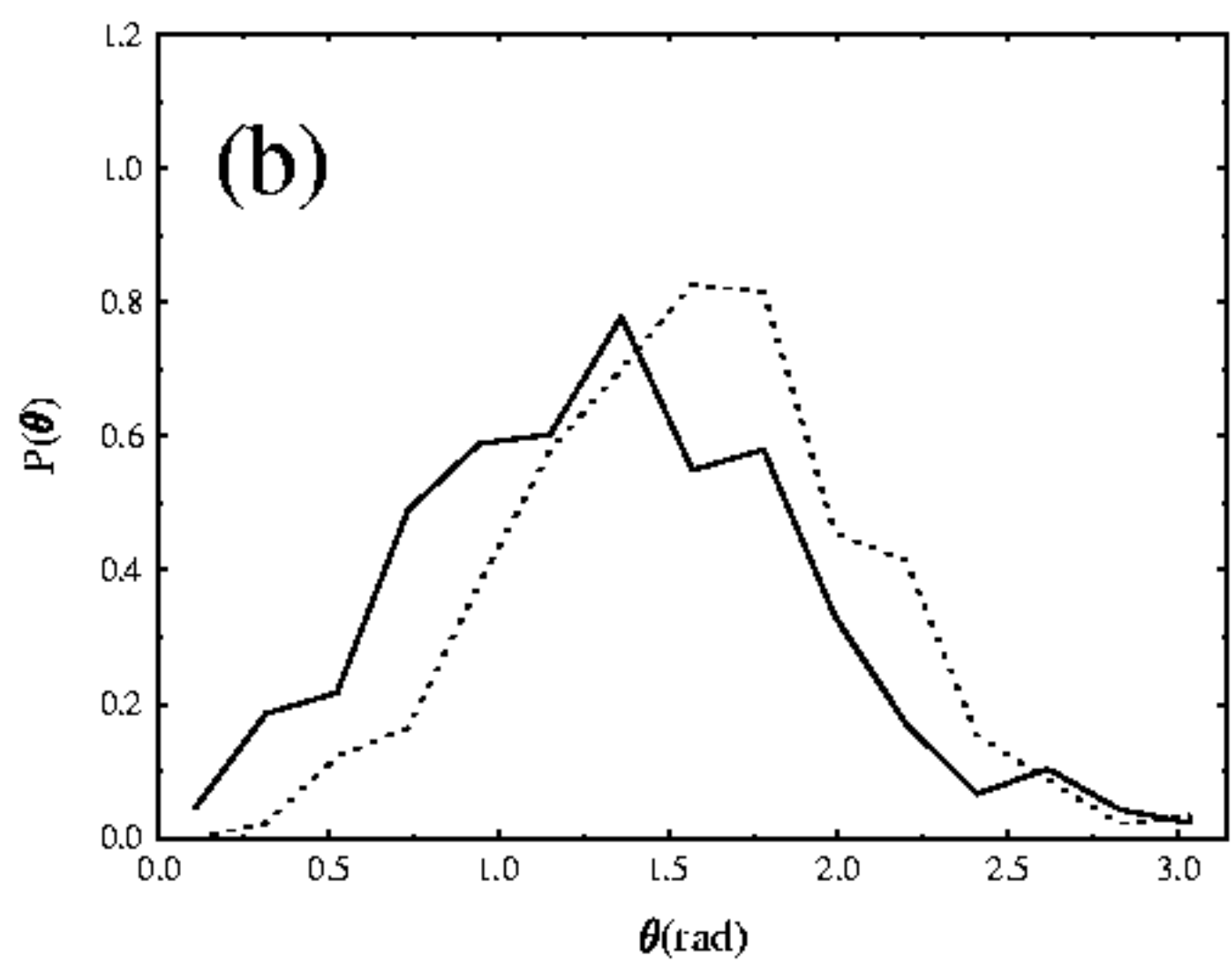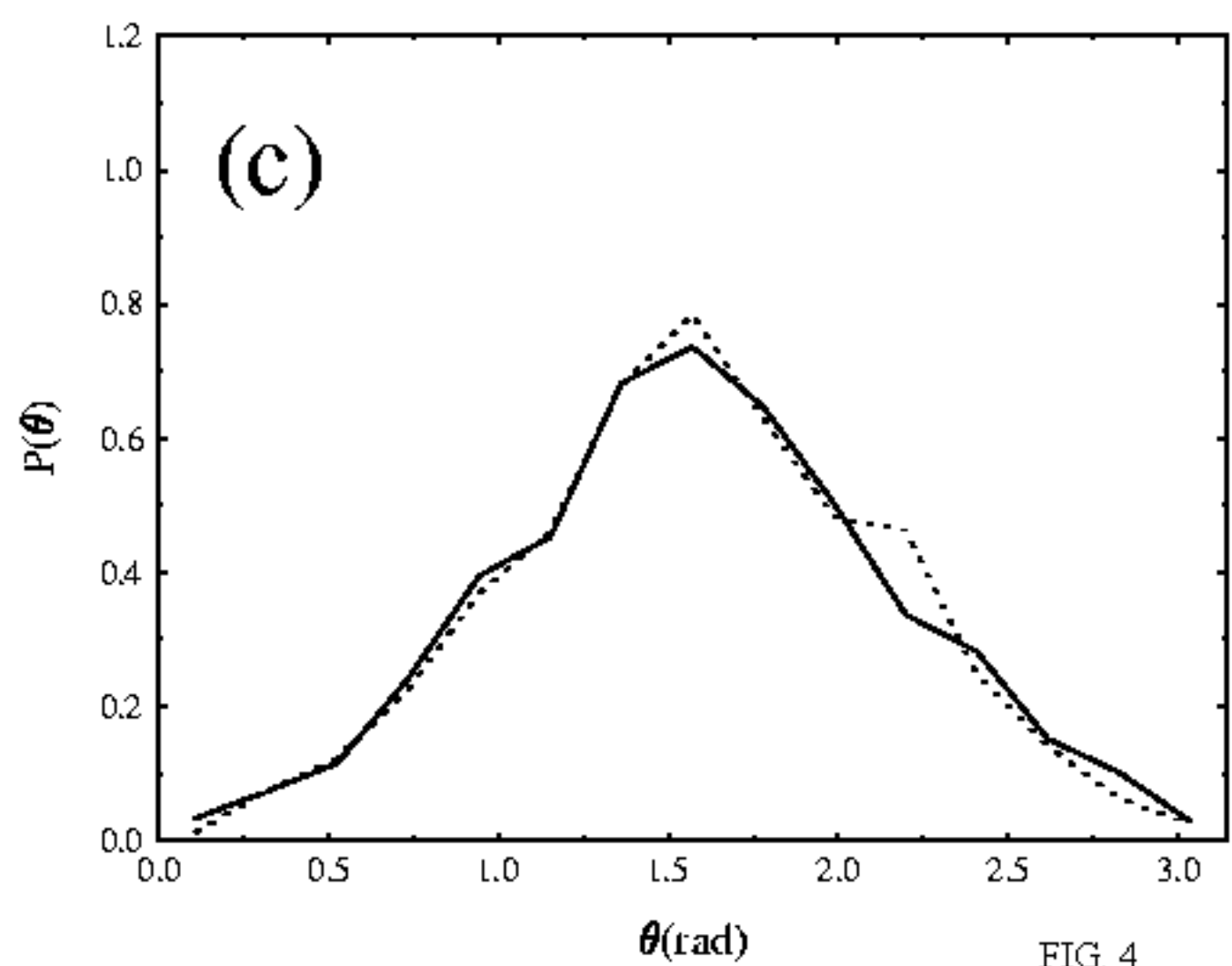

FIG. 4

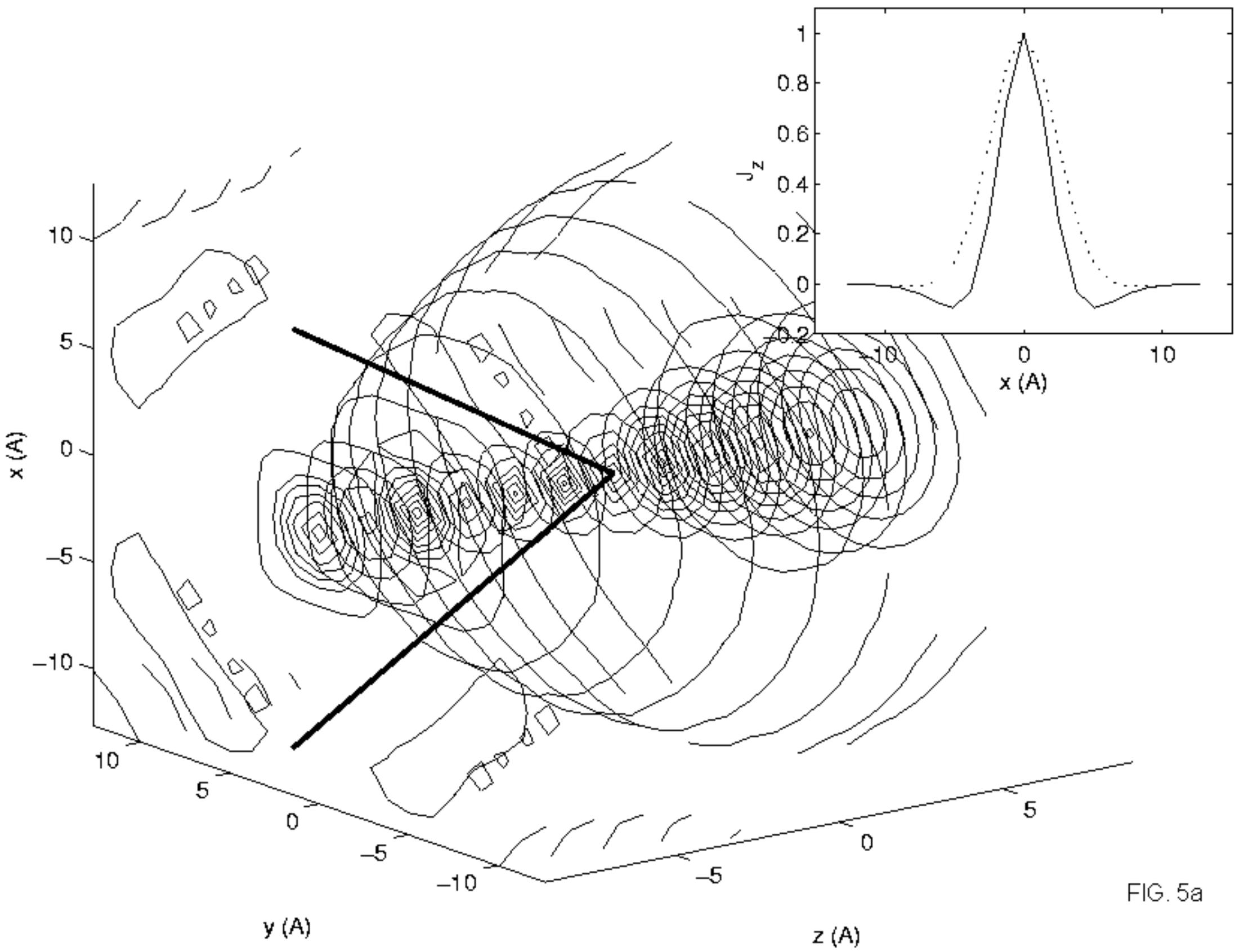

FIG. 5a

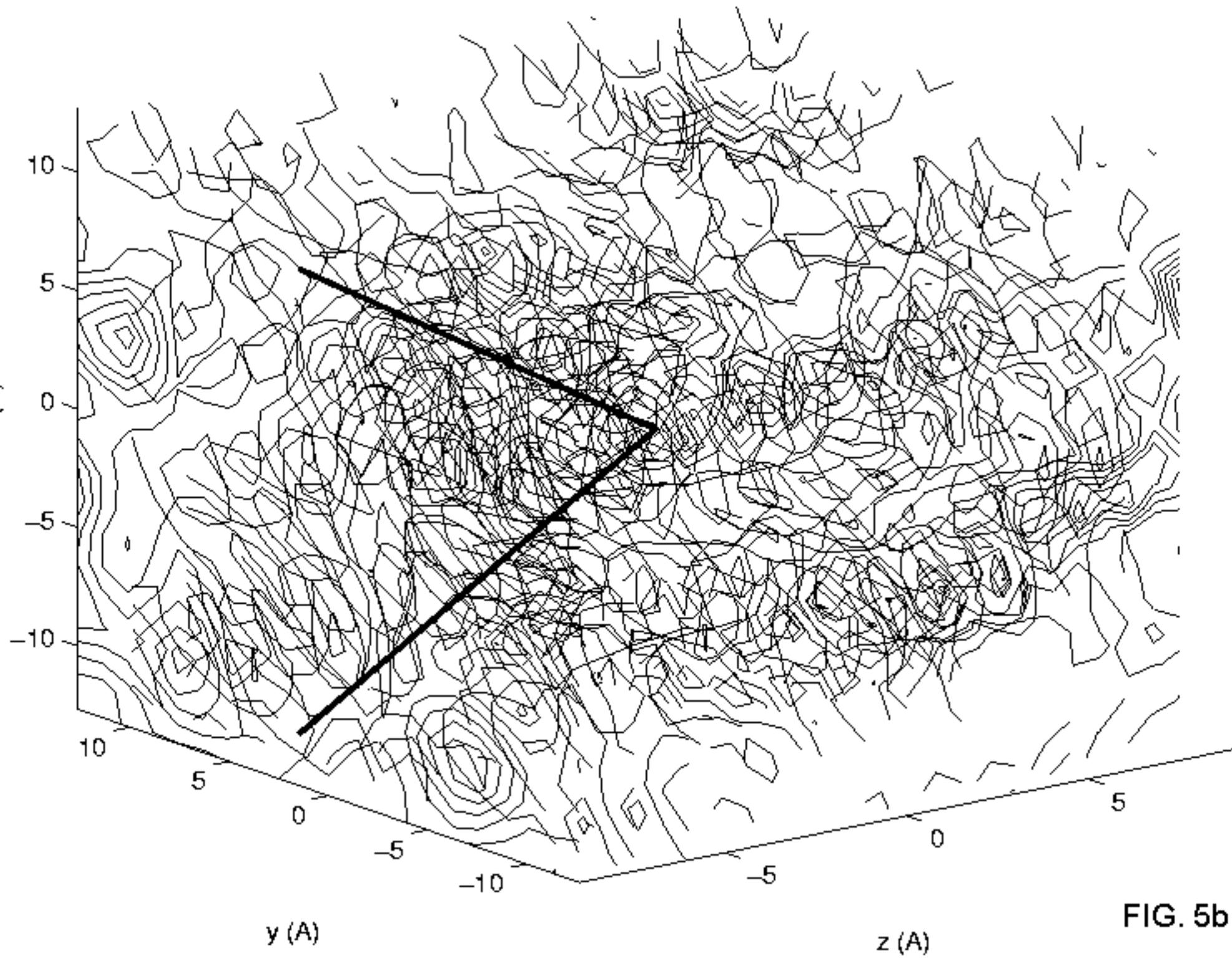

FIG. 5b

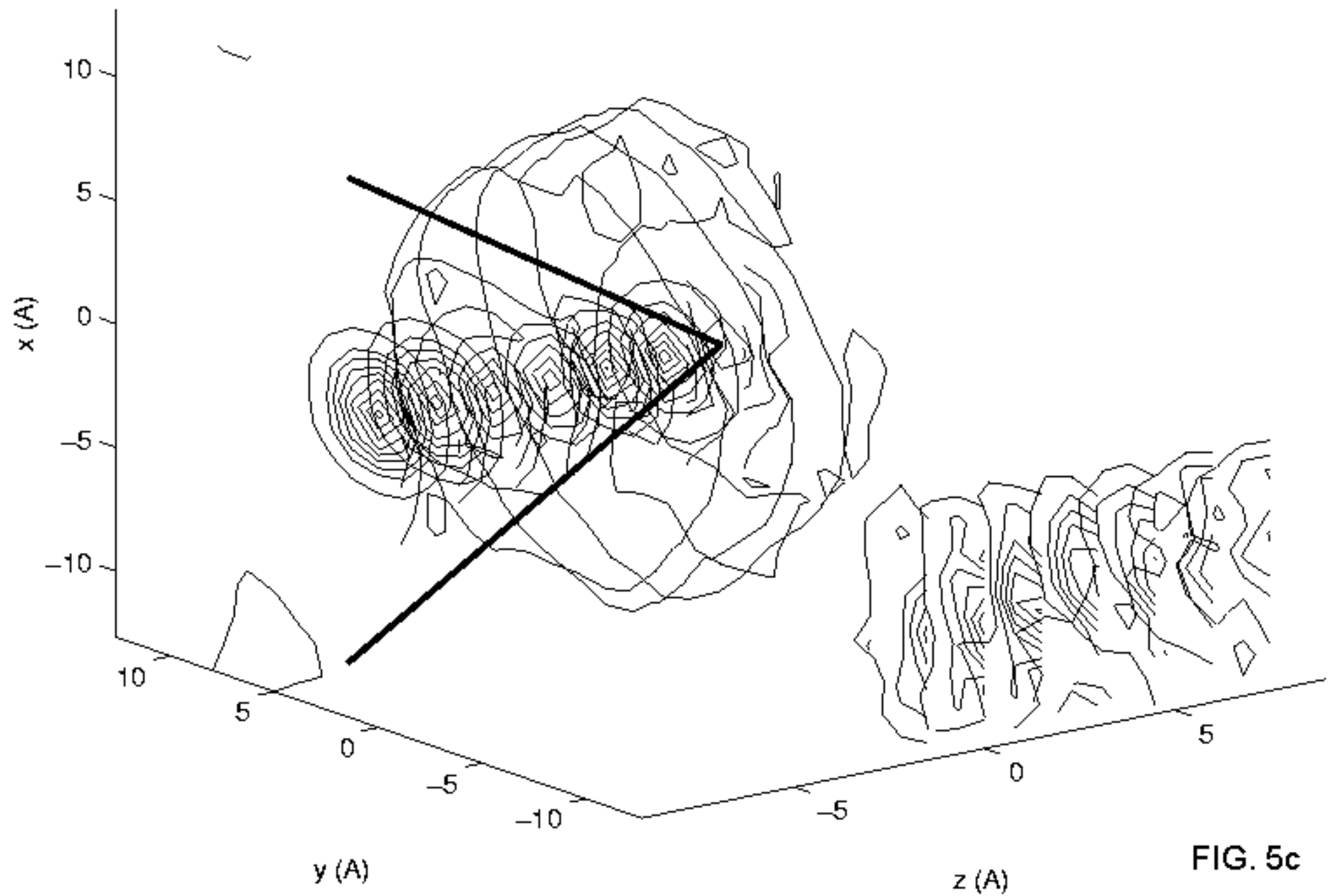

FIG. 5c

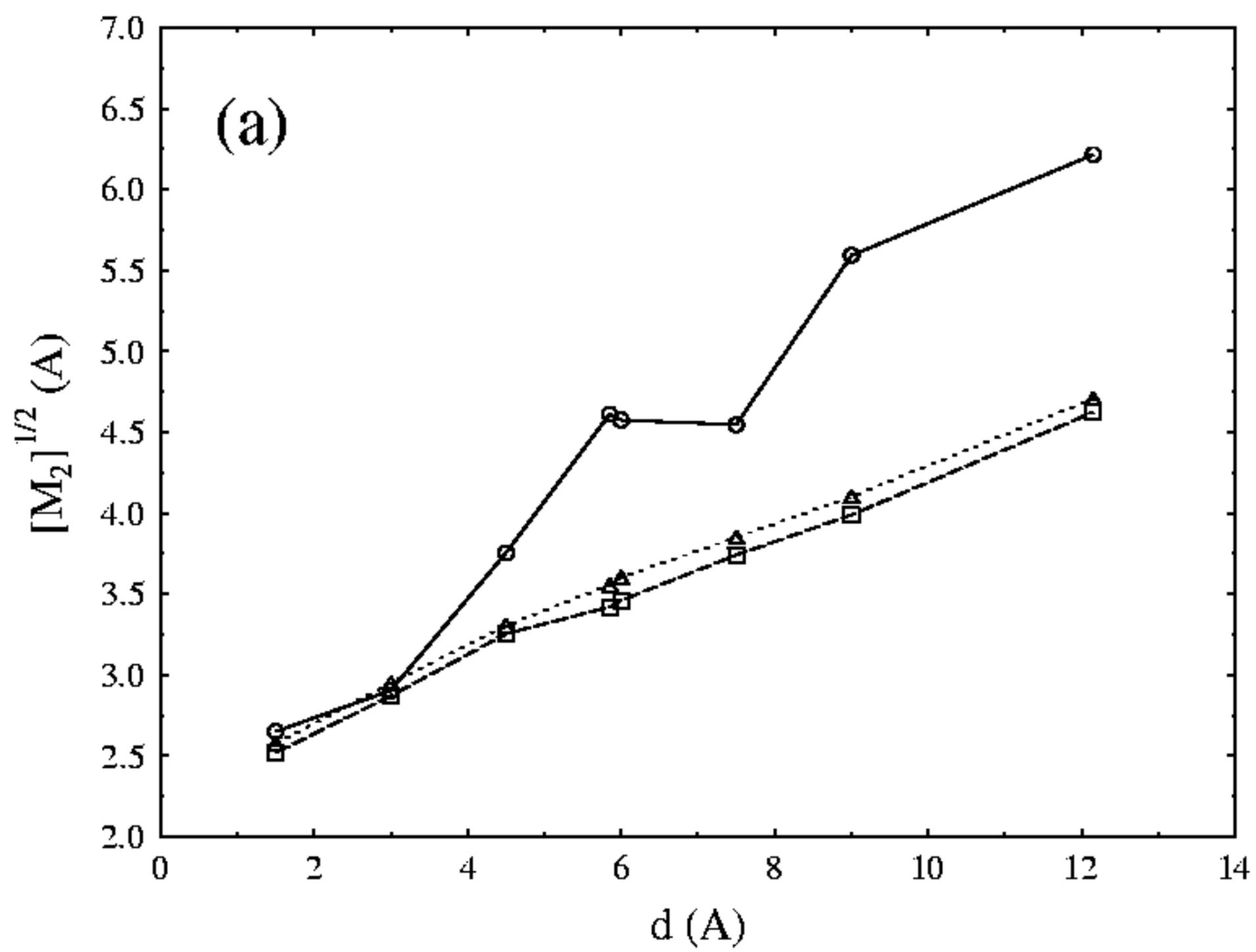
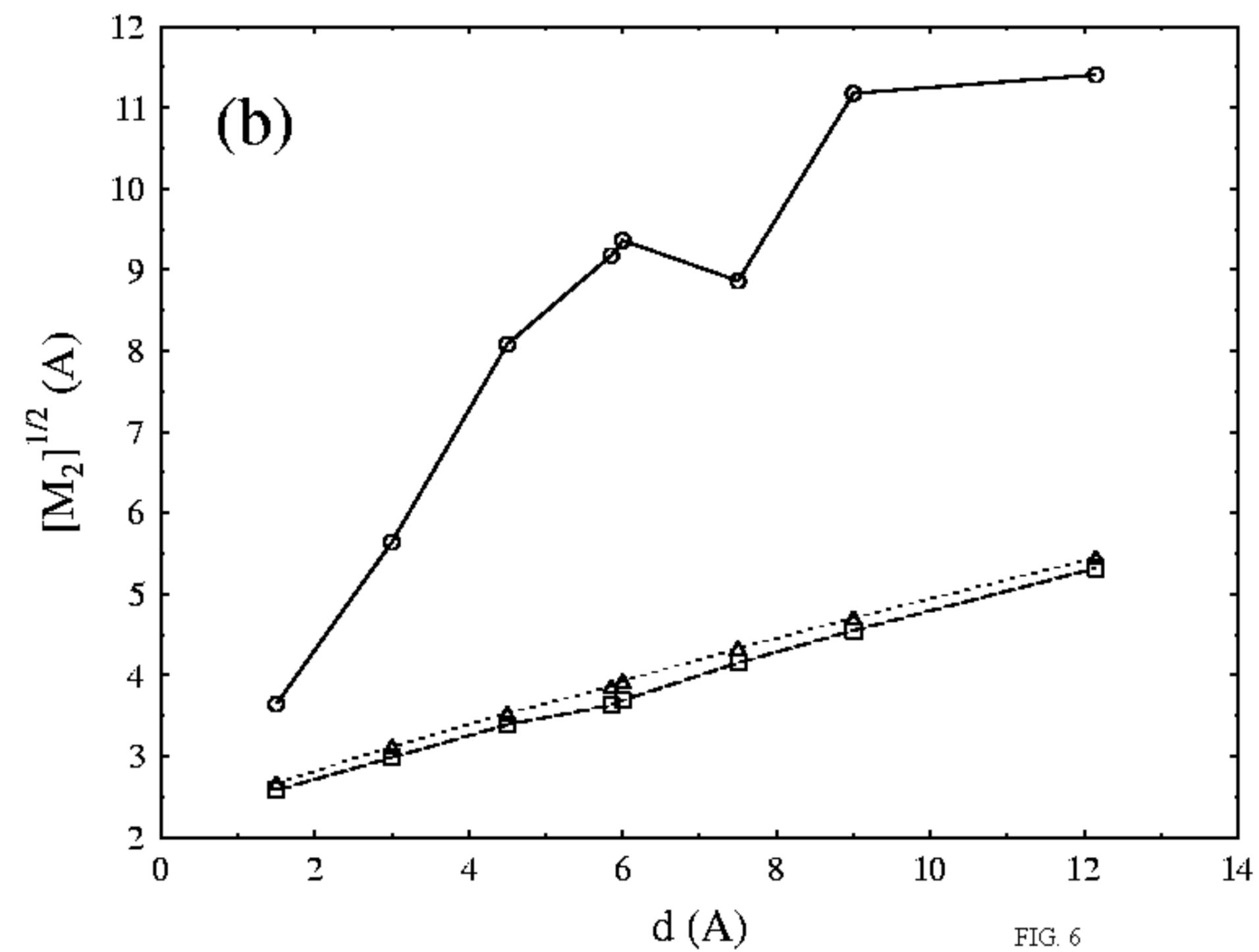

FIG. 6

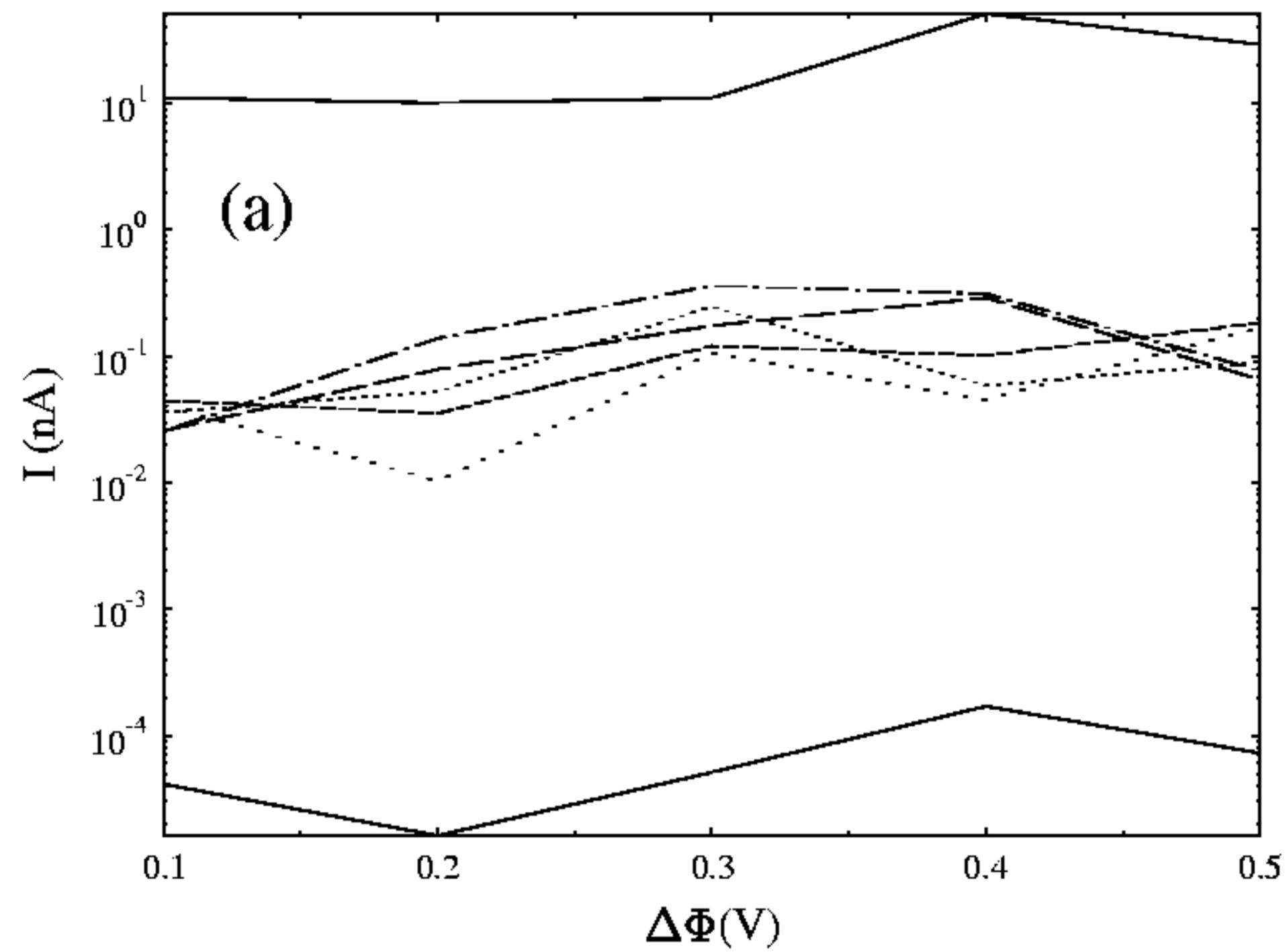

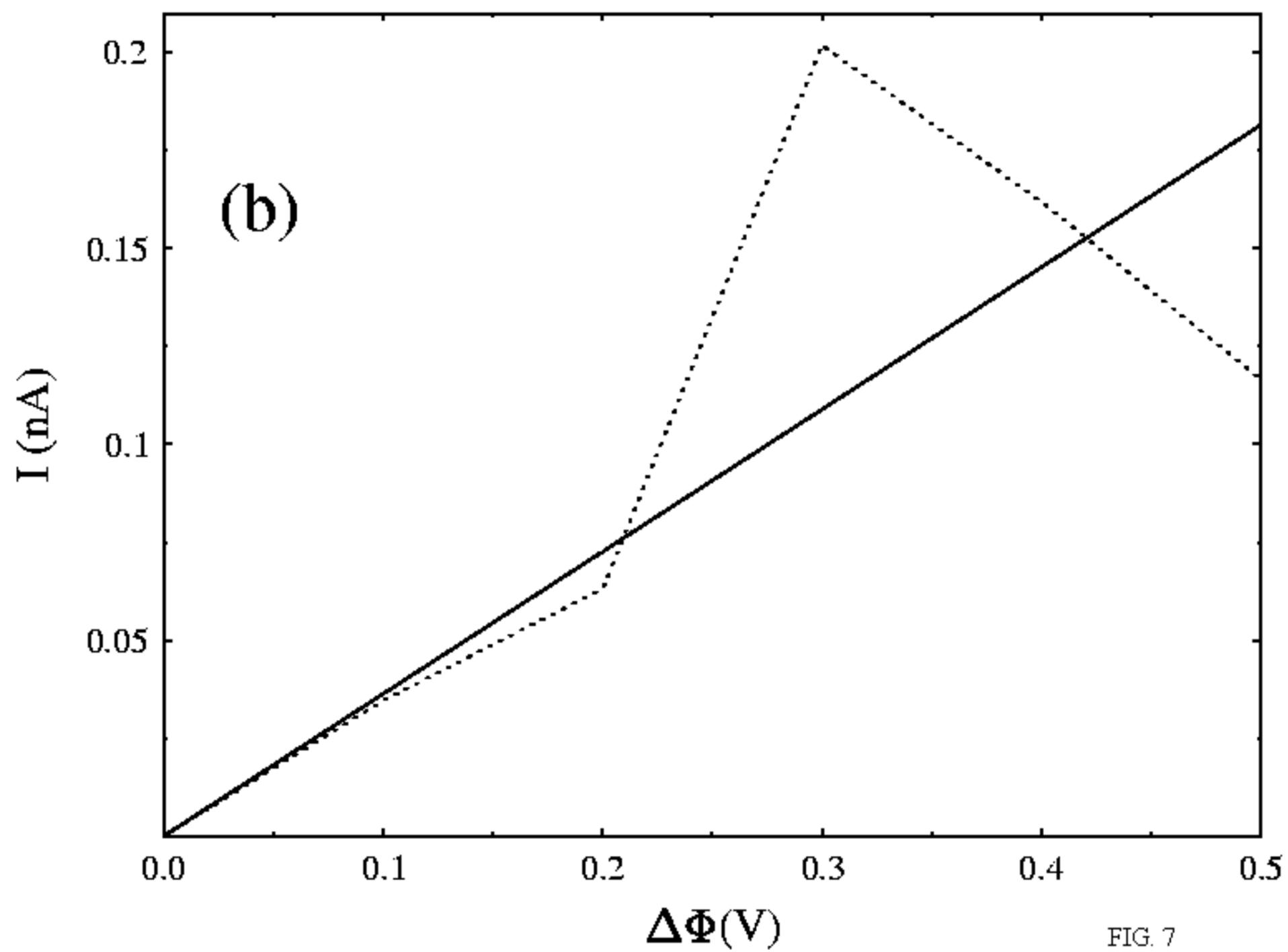

FIG. 7